\documentclass[amssymb,amsmath,prb,twocolumn,floatfix,showpacs,superscriptaddress]{revtex4}
\usepackage{graphicx}

\begin{document}
\title{Charge shelving and bias spectroscopy for the readout of 
a charge qubit on the basis of superposition states.}

\author{Andrew D. Greentre}

\affiliation{Centre for Quantum Computer Technology, School of Physics, The University of New South Wales, Sydney, NSW 2052, Australia}

\affiliation{Centre for Quantum Computer Technology, School of Physics, The University of Melbourne, Melbourne, Victoria 3010, Australia}

\author{A. R. Hamilton}
\affiliation{Centre for Quantum Computer Technology, School of Physics, The University of New South Wales, Sydney, NSW 2052, Australia}

\author{F. Green}
\affiliation{Centre for Quantum Computer Technology, School of Physics, The University of New South Wales, Sydney, NSW 2052, Australia}

\date{16 July 2004}

\begin{abstract}
\footnotesize{Charge-based qubits have been proposed as fundamental elements for
quantum computers.  One commonly proposed readout device is the
single-electron transistor (SET).  SETs can distinguish between
localized charge states, but lack the sensitivity to directly
distinguish superposition states, which have greatly enhanced
coherence times compared with position states.  We propose
introducing a third dot, and exploiting energy dependent
tunnelling from the qubit into this dot (bias spectroscopy) for
pseudo-spin to charge conversion and superposition basis readout.
We introduce an adiabatic fast passage-style charge pumping
technique which enables efficient and robust readout via charge
shelving, avoiding problems due to finite SET measurement time.}
\end{abstract}


\maketitle


The experimental observation, manipulation and utilization of
\textit{coherent} quantum mechanical properties in solid-state
systems are key technological challenges for this century.  The
importance of incoherent quantum properties has been essential for
the development of microelectronics and it is hoped that coherent
quantum effects will spawn new technologies including, but not
necessarily limited to, quantum computers \cite{bib:NielsenBook}.

In the development of coherent solid-state systems compatible with
quantum computing, superconducting systems have a clear advantage
due to the presence of macroscopic quantum states and key
milestones have already been  reached
\cite{bib:NakamuraNature1999,bib:VionScience2002,bib:PashkinNature2003}.
Coherent transport in semiconductor two-dimensional electron-gas
(2DEG) systems has been observed, \cite{bib:vanderWielRMP2003} and
recently  a charge qubit has been realized in a GaAs double dot
\cite{bib:HayashiPRL2003}.  Despite this, there is a strong
impetus to develop coherent technologies that are compatible with
the semiconductor industry, especially those based on
silicon-metal-oxide technology, owing to its mature manufacturing
technology and potential scalability advantages
\cite{bib:Kane,bib:ClarkPRS2003,bib:Hollenberg2003}.

Of particular interest are charge-based quantum computers
\cite{bib:EkertRMP1996}, for example in Cooper-pair box
arrangements \cite{bib:NakamuraNature1999,bib:MakhlinNature1999},
and semiconductor systems
\cite{bib:ClarkPRS2003,bib:Hollenberg2003,bib:Schirmer2003},
because of the relative ease of readout using high sensitivity
electrometers.  One such electrometer is the radio-frequency
single-electron transistor (rf-SET)
\cite{bib:SchoelkopfScience1998}, which has been shown to be
compatible with quantum computing requirements
\cite{bib:BuehlerC-M0304384}.  The relative ease of coupling to a
charge-based qubit is, however, also responsible for giving the
qubit a short decoherence time, as the charge distribution of the
qubit couples readily to the local electrostatic environment.  By
operating a charge qubit at the degeneracy point, the natural
basis is the superposition basis. These states couple less
strongly to the electrostatic environment, so the decoherence rate
of a superposition basis qubit should be much less than that for a
position basis qubit.  Conventional electrometers lack the
sensitivity to directly distinguish between superposition states.
Our scheme provides a robust mechanism to convert information from
the superposition basis to an accessible position basis,
incorporating charge shelving in a fashion not available with
conventional double-dot schemes.

\begin{figure}[b]
\includegraphics[height = 4.0cm,clip]{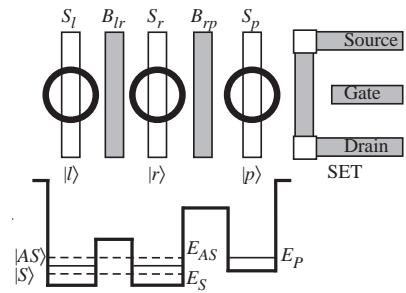}
\caption{\label{fig:TripleWell} \footnotesize{(Top) configuration showing
phosphorus donors and gate structure.  There is one electron and
an SET for readout.  The qubit is defined by donors marked
$|l\rangle$ and $|r\rangle$, monitoring the third donor,
$|p\rangle$ (probe) provides the readout.  State energies ($E$)
are controlled via gates $S_l$, $S_r$, and $S_p$, and the coherent
tunnelling rates, $\Omega_{lr}$ and $\Omega_{rp}$ are controlled
by barrier gates $B_{lr}$ and $B_{rp}$ respectively. (Bottom)
triple well diagram with $E_l=E_r$, the symmetric $|S\rangle$ and
antisymmetric $|AS\rangle$ states are equally separated from
$E_l$. $E_p$ has been tuned to $E_{AS}$ so resonant tunnelling
between $|AS\rangle$ and $|p\rangle$ occurs.}}
\end{figure}

We consider a coherent triple-dot, one-electron system, with a
strongly-coupled qubit, and a weakly-coupled `probe' dot,
illustrated in Fig. \ref{fig:TripleWell}.  Varying the energy of
the third dot, relative to the qubit, achieves a form of bias
spectroscopy, reminiscent of the optical Autler-Townes (AT)
experiment \cite{bib:ATExpt}. Such spectroscopy serves to probe
the system dynamics.  However for single-shot qubit readout we
propose an adiabatic fast passage (AFP) \cite{bib:VitanovARPC2001}
like process to perform superposition-to-position state pumping.
This constitutes a form of charge-shelving and is analogous to a
scheme for adiabatic transport in a double-dot system by Brandes
and Vorrath \cite{bib:BrandesPRB2002}. Although our scheme should
be applicable to any three-state, one-electron system, for clarity
we focus on the phosphorus-in-silicon system of Hollenberg
\textit{et al.} \cite{bib:Hollenberg2003}.  We note that subsequent to our initial suggestion, a similar scheme has been realized in a beautiful experiment by Astafiev \textit{et al.} \cite{bib:AstafievCM0402619} for a Josephson charge qubit.

Our scheme requires a small increase in the complexity required to
perform a charge-qubit measurement over simpler two-donor schemes.
Therefore we must identify the circumstances where it will be
advantageous.

For any quantum computing scheme there are many important
timescales, and the choice of an effective measurement scheme
depends on the relative values of each. There are the two
environmental dephasing times, $T_1$ and $T_2$ corresponding to
population and coherence relaxation respectively.
$\tau_{\mathrm{meas}}$ is the time required for a measurement to
occur, and because a measurement projects the system into a basis
state of the measurement device, this contributes an effective
$T_2$.  We define $\tau_{\mathrm{osc}}$ as the time for one
coherent oscillation to occur and $\tau_{\mathrm{gate}}^{-1}$ as
the maximum rate at which signals can be sent to manipulate the
qubit. For a functioning qubit we must have
$\tau_{\mathrm{osc}},\tau_{\mathrm{gate}} \ll T_1, T_2$, and this
is assumed in our discussion.

Nondestructive single-shot readout requires $\tau_{\mathrm{meas}}
\ll T_1$.  $T_1$ is a function of the energy separation between
states, and for good charge localization (necessary for SET
readout) we must operate far from degeneracy, i.e. where $T_1$ is
minimized.  Recent experiments and analysis of rf-SETs
\cite{bib:BuehlerC-M0304384} suggest that if the induced SET
island charge is $\sim 0.01 \mathrm{e}$, it will take
$\tau_{\mathrm{meas}} \sim 2 \mu\mathrm{s}$ to achieve an error
rate of $0.1$. This $\tau_{\mathrm{meas}}$ means that although
proof of principle experiments using signal averaging will be
possible, single shot measurements for readout and error
correction in a practical device will be problematic as the
population will decay faster than it can be measured.  By
introducing charge shelving to isolate the charge at the bias
position that \emph{maximizes} $T_1$ during readout, we ameliorate
this.

To exploit SET readout and superposition basis operation, we must
transfer quantum information from the superposition basis, to a
position basis.  This may be done either nonadiabatically or
adiabatically.

Nonadiabatic operations, which could be performed in either double
or triple dot geometries, require $\tau_{\mathrm{gate}} \ll T_1,
T_2$ and $\tau_{\mathrm{osc}}$.  If sufficient bandwidth is
available they constitute the fastest mechanism for transferring
population.  Nonadiabatic operations are extremely sensitive to
noise and gate errors, and hence attention has turned to adiabatic
methods.

There are at least two adiabatic timescales, the first being the
usual quasi-static case where $\tau_{\mathrm{gate}} \gg T_1, T_2,
\tau_{\mathrm{osc}}$.  Adiabatic transfer based on quasi-static
operations is conceptually easy and applicable to both double and
triple dot schemes.  By necessity such manipulations are slow, and
there may be incompatibility between $\tau_{gate}$ and $T_1$,
meaning such rotations are not suitable for quantum computer
readout. Triple-dot systems afford the possibility for a further
adiabatic timescale, where adiabatic \emph{passage} techniques
exploiting appropriate parameter modulation, allow decoupling of
the adiabatic pathway from the dephasing times
\cite{bib:VitanovARPC2001}, i.e. $\tau_{\mathrm{osc}} \ll
\tau_{\mathrm{gate}} \ll T_1, T_2$. It is this second adiabatic
timescale that we exploit for AFP charge-shelving.  For a more
complete discussion of timescales in coherently driven systems see
Ref. \cite{bib:GreentreePRA2002}.

\begin{figure}[tb]
\includegraphics[height=7.5cm,clip]{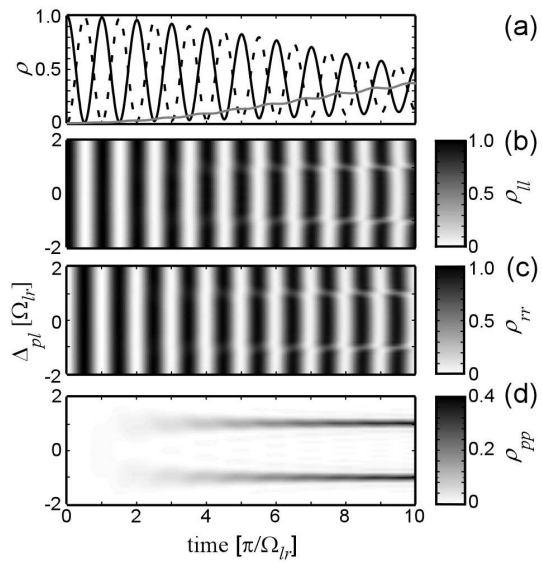}
\caption{\label{fig:TransientResPCs} \footnotesize{(a) $\rho_{ll}$ (black solid
line), $\rho_{rr}$ (black broken line) and $\rho_{pp}$ (grey solid
line) as a function of time (in units of $\pi/\Omega_{lr}$) for
$\Delta_{pl}=\Omega_{lr}=2$. Density plots showing $\rho_{ll}$
(b), $\rho_{rr}$ (c), and $\rho_{pp}$ (d), as a function of time
and $\Delta_{pl}$ (in units of $\Omega_{lr}$) for $E_l = E_r = 0$,
$\Omega_{rp} = \Omega_{lr}/20$ and $\Gamma = \Omega_{lr}/100$ for
a qubit in state $|l\rangle$ at time $t=0$.  Note the dominant
oscillatory behavior in $\rho_{ll}(t)$ and $\rho_{rr}(t)$ and the
AT doublet-like feature in $\rho_{pp}$.}}
\end{figure}

To summarize, any practical qubit requires
$\tau_{\mathrm{osc}},\tau_{\mathrm{gate}},\tau_{\mathrm{meas}} \ll
T_2, T_1$, and most implementations proposed to date also require
$\tau_{\mathrm{gate}} \ll \tau_{\mathrm{osc}}$.  Our scheme can
function with $\tau_{\mathrm{gate}} > \tau_{\mathrm{osc}}$, to
identify the signature of coherent oscillations, and
charge-shelving scheme provides a mechanism to increase $T_1$ to
ensure that $\tau_{\mathrm{meas}} \ll T_1$.  AFP provides the
advantages of adiabatic control without restricting gate
operations to the timescales for decoherence rates.  Even so, it
is necessary to explore any given implementation fully to
determine if our scheme will provide a tangible advantage.

One further consideration is that access to multiple bases is
necessary for state tomography \cite{bib:JamesPRA2001}, which is
important for qubit characterization.   The extra freedom afforded
by the triple-donor system suggests that a hybrid
position/superposition readout system may be realizable which
would have advantages for tomography and we will investigate this
possibility elsewhere.

The three-donor system is shown in Fig. \ref{fig:TripleWell}, with
three ionized phosphorus donors (open circles) sharing a single
electron.   A strongly coupled qubit is defined by donors $l$
(left) and $r$ (right).  The weakly coupled probe is labelled $p$.
We follow the gate notation used in Refs.
\cite{bib:Hollenberg2003,bib:Schirmer2003}. The energies of each
single-electron state are controlled using shift gates (open
rectangles), $S$, and the energies of these states are $E_l$,
$E_r$, $E_p$, with $\Delta_{\alpha,\beta} = E_{\alpha}-E_{\beta}$,
$\alpha,\beta = l,r,p$. Coherent tunnelling is controlled by
barrier gates (grey rectangles), which vary the barrier height.
The barrier gate between donors $l$ and $r$ ($r$ and $p$) is
labelled $B_{lr}$ ($B_{rp}$). We ignore direct tunnelling between
$l$ and $p$.  An SET reads out the electron on $p$.  The natural
basis for the qubit will be the superposition basis, we write the
symmetric (anti-symmetric) state as $|S\rangle = (1/\sqrt{2})
(|l\rangle + |r\rangle$) ($|AS\rangle = (1/\sqrt{2}) (|l\rangle -
|r\rangle)$).

\begin{figure}[tb]
\includegraphics[height=7.5cm,clip]{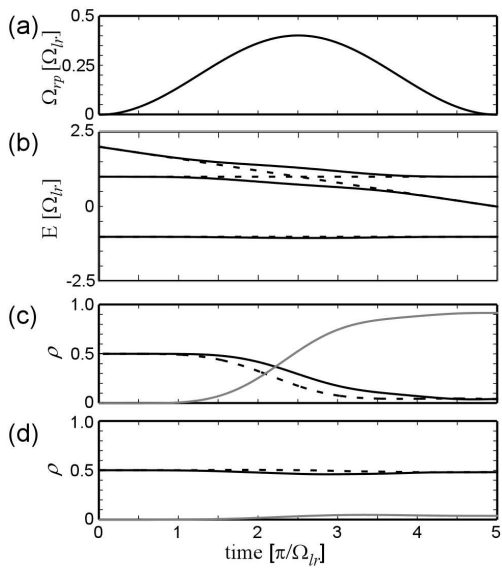}
\caption{\label{fig:AFP} \footnotesize{AFP readout (a) $\Omega_{rp}$ as a
function of time (in units of $\pi/ \Omega_{lr}$) showing
modulation of the coherent tunnelling rate. (b) State energies (in
units of $\Omega_{lr}$) as a function of time.  The dashed, black
lines represent the energies unperturbed by $\Omega_{rp}$, whilst
the solid blue lines represent state energies with
$\Omega_{rp}(t)$ applied. At $t=0$ the states are (highest to
lowest energy) $|p\rangle$, $|AS\rangle$ and $|S\rangle$. (c),(d)
$\rho_{ll}$ (black solid line), $\rho_{rr}$ (black broken line)
and $\rho_{pp}$ (grey solid line) as a function of time for
$\rho(0) = |AS\rangle$ (c) $\rho(0) = |S\rangle$ (d).  In all
cases $\Omega_{rp}^{\max} = 0.4\Omega_{lr}$.}}
\end{figure}

We vary $E_p$ and monitor the population in $p$ with the SET,
observing nonzero populations in $p$ only when $E_p$ is
quasi-degenerate with either of the superposition states.  Similar
bias spectroscopy is seen in open 2DEG systems
\cite{bib:vanderWielRMP2003}.  In optical AT experiments, it is
more usual to monitor the response of a weak probe field, which is
proportional to the coherence $\rho_{rp}$.

We solve the density matrix equations of motion to give both
transient and spectroscopic insight into the dynamics.  This is
similar to the approach in Ref. \cite{bib:GreentreePRA2002}.  We
express the coherent tunnelling rate (analogous to the Rabi
frequency in optical experiments) as $\Omega_{\alpha\beta}=2 \pi /
\tau_{\mathrm{osc}}^{\alpha \beta}$, where the $B$ gate dependance
has been suppressed. This follows the spirit of Gurvitz's
\cite{bib:GurvitzPRB1997} treatment for the two dot system and
that of Renzoni and Brandes \cite{bib:RenzoniPRB2001} in a
triple-well system. The Hamiltonian is
\begin{eqnarray*}
{\cal H} &=& \Delta_{rl} |r\rangle \langle r|
          + \Delta_{pl} |p\rangle \langle p|
\nonumber \\
          & & - \hbar\Omega_{lr}
          (|l\rangle \langle r| + |r\rangle \langle l|)
          - \hbar\Omega_{rp}
          (|r\rangle \langle p| + |p\rangle \langle r|).
\end{eqnarray*}
This Hamiltonian is identical to an optically driven three-level
atom in the rotating-wave approximation \cite{bib:ShoreBooks}.

The density matrix equations of motion are written
\begin{eqnarray*}
\dot{\rho} = -(i / \hbar)[{\cal H},\rho] + {\cal L},
\end{eqnarray*}
with $\rho$ the density matrix and ${\cal L}$ a dephasing
operator.  $T_2$ processes are modelled by a dephasing rate
$\Gamma$ which is assumed affect all coherences, $T_1$ processes
are described by rates of incoherent population transfer
\cite{bib:TothPRB1999}
\begin{eqnarray*}
\Gamma_{\alpha\beta}=
\chi_{\alpha\beta}\Delta_{\alpha\beta}/\left[1-\exp(\Delta_{\alpha\beta}/kT)\right],
\end{eqnarray*}
where $k$ is Boltzmann's constant, $T$ the temperature and
$\chi_{\alpha\beta}$ a rate from the tunnelling probability
between $\alpha$ and $\beta$. $\Gamma_{\alpha\beta}$ is the
population transfer rate from $\beta$ to $\alpha$.

The density matrix equations of motion are therefore
\begin{eqnarray}
\dot{\rho}_{ll} &=& i \Omega _{lr}
    (\rho_{rl}-\rho _{lr})
    -\Gamma_{rl} \rho_{ll} + \Gamma_{lr} \rho_{rr},
\nonumber \\
\dot{\rho}_{lr} &=& i
    \left[\frac{\Delta_{rl}}{\hbar} \rho _{lr}
    + \Omega _{lr} (\rho _{rr} - \rho _{ll})
    - \Omega_{rp}\rho_{lp} \right]
    -\Gamma \rho_{lr},
\nonumber \\
\dot{\rho}_{lp} &=& i
    \left[\frac{\Delta_{pl}}{\hbar} \rho_{lp}
    + \Omega_{lr} \rho _{rp}
    - \Omega_{rp} \rho _{lr}\right]
    - \Gamma \rho_{lp},
\nonumber \\
\dot{\rho}_{rr} &=& i
    \left[\Omega _{lr} (\rho_{lr}-\rho_{rl})
    + \Omega_{rp} (\rho_{pr}-\rho_{rp})\right]
\nonumber \\
    & &+ \Gamma_{rl} \rho_{ll}
    - (\Gamma_{lr} + \Gamma_{pr}) \rho_{rr}
    + \Gamma_{rp} \rho_{pp},
\nonumber \\
\dot{\rho}_{rp} &=& i
    \left[\frac{\Delta_{pr}}{\hbar} \rho_{rp}
    + \Omega_{rp} (\rho_{pp}-\rho_{rr})
    + \Omega_{lr}\rho_{lp} \right]
    - \Gamma \rho_{rp},
\nonumber \\
\dot{\rho}_{pp} &=& i
    \Omega_{rp}(\rho_{rp}-\rho_{pr})
    + \Gamma_{pr}\rho_{rr}
    - \Gamma_{rp} \rho_{pp}, \nonumber \\
\rho _{\alpha \beta } &=& \rho _{\beta \alpha }^{\ast }  \nonumber \\
1 &=& \rho_{ll}+\rho_{rr}+\rho_{pp}. \label{eq:DensMatClosed}
\end{eqnarray}


We numerically integrated Eqs. \ref{eq:DensMatClosed} with
$\rho_{ll}(0) = 1$ to highlight the dynamics. Maximum coherence
times require $E_l=E_r$ which is used in our calculations. For the
third dot to act as a weak probe, $\Omega_{rp} \ll \Omega_{lr}$.
We have therefore chosen $\Gamma = \Omega_{lr}/100$,
$\chi_{\alpha\beta} = 0$ and $\Omega_{rp}=\Omega_{lr}/20$.  Fig.
\ref{fig:TransientResPCs} (a) shows the time dependent populations
when $E_p=E_{AS}$.  The dominant feature is the coherent
population oscillations between $l$ and $r$.  There is a steady
buildup of population in $p$, which is our measurement signal.
Figs. \ref{fig:TransientResPCs} (b)-(d) show $\rho_{ll}$,
$\rho_{rr}$, and $\rho_{pp}$ respectively, as a function of time
and $\Delta_{pr}$.  Again, the dominant behavior is the coherent
oscillation between $l$ and $r$, however when $\Delta_{pr}=\pm
\Omega_{lr}$, resonant tunnelling into $p$ occurs, yielding a
doublet in $\rho_{pp}$ similar to the AT doublet.  Measurement of
probe population is therefore sensitive to the population in the
symmetric and anti-symmetric modes of the $l-r$ system.


Bias spectroscopy should be useful for characterizing qubit
properties, albeit incompatible with readout of a practical
quantum computer. This is due to (i) coherent oscillations on the
$r-p$ transition reducing readout fidelity, (ii) the need to set
and accurately maintain $\Delta_{pl}$ over the measurement time,
and (iii) small average populations tunnelling into $p$ requiring
multiple experiments.  To avoid these problems, we propose a form
of charge-shelving that adiabatically drives the population into
$|p\rangle$ from one of the superposition states, realized by
controlling the tunnelling in a fashion related to AFP. The
advantages of this are that $\rho_{pp}$ is adiabatically
\textit{driven} to a large value (approaching unity) in a short
time (typically a few $\Omega_{lr}^{-1}$), with robustness to gate
errors. Because of the energy dependence of $T_1$, it is most
useful to perform AFP between state $|p\rangle$ and the more
energetic of the two superposition states, i.e. $|AS\rangle$. One
must take care in choosing the modulation trajectory for
$|p\rangle$ in this case, as there will be some off-resonant
interactions. Our results are promising however, and further
optimization can be done.

The trajectory taken by state $|p\rangle$ is governed by both
$\Delta_{pl}$ and $\Omega_{rp}$ and for the traces in Figs.
\ref{fig:AFP} they were:
\begin{eqnarray*}
    \Delta_{lp} &=& 2\Omega_{lr}\left(1-t/t_{\max}\right), \\
    \Omega_{rp} &=& \Omega_{rp}^{\max}
    \left[ 1 - \cos\left(2\pi t / t_{\max} \right) \right]/2,
\end{eqnarray*}
where $\Omega_{rp}^{\max} = 0.4\Omega_{lr}$, $t_{\min}=0$ and
$t_{\max} = 5 \pi/\Omega_{lr}$. In order to make this more
explicit, Fig. \ref{fig:AFP} (a) shows $\Omega_{rp}(t)$, in
keeping with conventional AFP schemes, our scheme is fairly
insensitive to the exact form of $\Omega_{rp}$; and (b) is a
diagram showing the energy levels as a function of time.

Fig. \ref{fig:AFP} (c) shows the populations for the qubit being
initially prepared in $|AS\rangle$.  After AFP most of the
population has been driven into $|p\rangle$.  Similarly Fig.
\ref{fig:AFP} (d) shows the effect of the AFP trajectory on
$|S\rangle$, in this case there is minimal population transfer. We
are presently performing more detailed numerical experiments to
optimize the population transfer and examine tomographic
applications of the scheme, as foreshadowed above. One important
issue is the potential re-initialization of the qubit after
readout. The AFP scheme as presented is entirely time reversible,
and therefore one can simply reverse the scheme to pump an
electron from the probe state into the anti-symmetric state to
reset the qubit.

In summary, we have presented a scheme for performing bias
spectroscopy on a qubit, where resonant tunnelling from the
superposition states to a third dot is read out with a SET.  Such
a scheme is a solid-state analog of the optical Autler-Townes
scheme. Using bias spectroscopy, one can map out the energy-level
space and this may prove a useful probe of coherent coupling where
conventional electrometers are unable to resolve the dynamics.
Because the bias spectroscopy described here should be able to
resolve arbitrary energy differences, this idea may be applied to
discriminating between the singlet and triplet states of a
two-spin system, such as in Ref. \cite{bib:Kane}.  This
constitutes an alternate approach to spin readout which we will
describe further elsewhere \cite{bib:GreentreeCM2004}. For useful readout, we propose an AFP
style scheme, where population is driven into the probe state.
This has the advantage of high fidelity readout with robustness to
gate errors. It also introduces a form of charge-shelving,
avoiding problems due to measurement times long compared with
$T_1$.

The authors would like to thank L. C. L. Hollenberg (University of
Melbourne) and H. S. Goan (University of Queensland) for useful
discussions. This work was supported by the Australian Research
Council, the Australian government and by the US National Security
Agency (NSA), Advanced Research and Development Activity (ARDA)
and the Army Research Office (ARO) under contract number
DAAD19-01-1-0653.



\begin{thebibliography}{99}
\footnotesize{
\bibitem{bib:NielsenBook} M. A. Nielsen and I. L. Chuang,
\textit{Quantum Computation and Quantum Information} (Cambridge
University Press, Cambridge, England 2000).

\bibitem{bib:NakamuraNature1999} Y. Nakamura, Yu. A. Pashkin, and
J. S. Tsai, Nature {\bf 398}, 786 (1999).

\bibitem{bib:VionScience2002} D. Vion, A. Aassime, A. Cottet, P. Joyez, H. Pothier, C. Urbina, D. Esteve, and M. H. Devoret, Science {\bf 296}, 886 (2002).

\bibitem{bib:PashkinNature2003} Yu. A. Pashkin, T. Yamamoto, O. Astafiev, Y. Nakamura, D. V. Averin, and J. S. Tsai, Nature {\bf 421}, 823 (2003).

\bibitem{bib:vanderWielRMP2003} W. G. van der Wiel, S. De Franceschi, J. M. Elzerman, T. Fujisawa, S. Tarucha, and L. P. Kouwenhoven , Rev. Mod. Phys. \textbf{75}, 1 (2003).

\bibitem{bib:HayashiPRL2003} T. Hayashi, T. Fujisawa, H. D. Cheong, Y. H. Jeong, and Y. Hirayama, Phys.
Rev. Lett. \textbf{91}, 226804 (2003).

\bibitem{bib:Kane} B. E. Kane, Nature {\bf 393}, 133 (1998); B. E. Kane, N. S. McAlpine, A. S. Dzurak, R. G. Clark, G. J. Milburn, H. B. Sun, and H. Wiseman , Phys. Rev. B {\bf 61}, 2961 (2000).

\bibitem{bib:ClarkPRS2003} R. G. Clark \textit{et al.}, Philos. Trans. R. Soc. Lond., Ser. A \textbf{361}, 1451 (2003).

\bibitem{bib:Hollenberg2003} L. C. L. Hollenberg, A. S. Dzurak, C. Wellard, A. R. Hamilton, D. J. Reilly, G. J. Milburn, and R. G. Clark, Phys. Rev. B \textbf{69}, 113301 (2004).

\bibitem{bib:EkertRMP1996} A. Ekert and R. Josza, Rev. Mod. Phys. \textbf{68}, 733 (1996).

\bibitem{bib:MakhlinNature1999} Yu. Makhlin, G. Sch\"{o}n, A. Shnirman, Nature \textbf{398}, 786 (1999).

\bibitem{bib:Schirmer2003} S. G. Schirmer, A. D. Greentree, and D. K.
L. Oi, quant-ph/0305052 (2003).

\bibitem{bib:SchoelkopfScience1998} R. J. Schoelkopf, P. Wahlgren, A. A. Kozhevnikov, P. Delsing, and D. E. Prober, Science \textbf{280}, 1238 (1998).

\bibitem{bib:BuehlerC-M0304384}  T. M. Buehler, D. J. Reilly, R. P. Starrett, A. D. Greentree, A. R. Hamilton, A. S. Dzurak, and R. G. Clark, cond-mat/0304384 (2003).

\bibitem{bib:ATExpt} S. H. Autler and C. H. Townes, Phys. Rev. {\bf 100} 703 (1955).

\bibitem{bib:VitanovARPC2001} N. V. Vitanov, T. Halfmann, B. W. Shore, and K. Bergmann, Annu. Rev. Phys. Chem. {\bf 52}, 763 (2001).

\bibitem{bib:BrandesPRB2002} T. Brandes and T. Vorrath, Phys. Rev.
B \textbf{66}, 075341 (2002).

\bibitem{bib:AstafievCM0402619} O. Astafiev, Yu. A. Pashkin, T. Yamamoto, Y. Nakamura, and J. S. Tsai, Phys. Rev. B \textbf{69}, 180507 (2004).

\bibitem{bib:GreentreePRA2002}     A. D. Greentree, T. B. Smith, S. R. de Echaniz, A. V. Durrant, J. P. Marangos, D. M. Segal, and J. A. Vaccaro, Phys. Rev. A, {\bf 65}, 053802 (2002).

\bibitem{bib:JamesPRA2001} D. F. V. James, P. G. Kwiat, W. J.
Munro, and A. G. White, Phys. Rev. A \textbf{64}, 052312 (2001).

\bibitem{bib:GurvitzPRB1997} S. A. Gurvitz, Phys. Rev. B {\bf 56},
15 215 (1997).

\bibitem{bib:RenzoniPRB2001} F. Renzoni and T. Brandes, Phys. Rev.
B {\bf 64}, 245301 (2001).

\bibitem{bib:ShoreBooks} B. W. Shore, \textit{The Theory of Coherent
Atomic Excitation} (Wiley, New York, 1990).

\bibitem{bib:TothPRB1999} G. T\'{o}th, A. O. Orlov, I. Amlani, C. S. Lent, G. H. Bernstein, and G. L. Snider, Phys. Rev. B \textbf{60}, 16906 (1999).

\bibitem{bib:GreentreeCM2004} A. D. Greentree, A. R. Hamilton, L. C. L. Hollenberg, and R. G. Clark, cond-mat/0403449 (2004).}

\end{thebibliography}
\end{document}